\documentclass[11pt]{article}

\pdfoutput=1

\usepackage{epsfig}
\usepackage{graphics}
\usepackage{epstopdf}
\usepackage{color}
\usepackage{tabularx,ragged2e,booktabs}
\usepackage{algorithmic}
\usepackage{algorithm}
\usepackage{fixltx2e}
\usepackage{setspace}
\usepackage[table]{xcolor}
\usepackage[english]{babel}
\usepackage{soul}
\usepackage{authblk}

\newcommand{\rem}[1]{}

\title{Parallel Distributed Breadth First Search on the Kepler Architecture}

\author[1]{Mauro Bisson\thanks{mauro.bis@gmail.com}}
\author[1]{Massimo Bernaschi}
\author[1]{Enrico Mastrostefano}

\affil[1]{Istituto per le Applicazioni del Calcolo, IAC-CNR, Rome, Italy}

\begin{document}
\date{}
\maketitle

\begin{abstract}
{\it
We present the results obtained by using an evolution of our CUDA-based
solution for the exploration, via a Breadth First Search, of large graphs. This
latest version exploits at its best the features of the Kepler architecture and
relies on a combination of techniques to reduce both the number of
communications among the GPUs and the amount of exchanged data. The final
result is a code that can visit more than 800 billion edges in a second by
using a cluster equipped with 4096 Tesla K20X GPUs.
}
\end{abstract}

\section{Introduction}

Many graph algorithms spend a large fraction of their time in accessing memory.
Moreover, on distributed memory systems, much time is spent in communications
among computational nodes.  A number of studies have focused on the Breadth
First Search (BFS) algorithm.  The BFS is a fundamental step for solving basic
problems on graphs, such as finding the diameter or testing a graph for
bi-partiteness, as well as more sophisticated ones, such as finding community
structures in networks or computing the maximum-flow/minimum-cut, problems that
may have an immediate practical utility.  Strategies for the optimization of
the BFS strongly depend on the properties of the analyzed graph. For example,
graphs with large diameters and regular structures, like those used in physical
simulations, present fewer difficulties compared to ``real-world" graphs that
can have short diameters and skewed degree distributions.  In the latter case,
a small fraction of vertices have a very high degree, whereas most have only a
few connections.  This characteristic complicates the load-balancing among
tasks in a parallel implementation.  Nevertheless, many authors have
demonstrated that it is possible to realize high performance parallel and
distributed BFS \cite{Agarwal:2010, Hong:2010, Merrill:2012, Beamer:2012,
Buluc:2011, Ueno:2012, Petrini2014}.

In the recent past we presented two codes \cite{Mastro2013, Bernaschi2014} able
to explore very large graphs by using a cluster of GPUs.  In the beginning, we
proposed a new method to map CUDA threads to data by means of a prefix-sum and
a binary search operation. Such mapping achieves a perfect load-balancing: at
each level of a Breadth First Search (BFS) one thread is associated with one
child of the vertices in the current BFS queue or, in other words, one thread
is associated to one vertex to visit.  Then we modified the Compressed
Sparse Row (CSR) data structure by adding a new array that allows for a faster
and complete filtering of already visited edges.  Moreover, we reduced the data
exchanged by sending the predecessors of the visited vertices only in the end
of the BFS.

In the present work we further extend our work in two directions: {\em i)} on
each single GPU we implement a new approach for the local part of the search that
relies on the efficiency of the atomic operations available in the Nvidia Kepler
architecture; {\em ii)} for the multi-GPU version, we follow a different
approach for the partitioning of the graph among GPUs that is based on a 2D
decomposition \cite{Yoo:2005} of the adjacency matrix that represents the graph.
The latter change improves the scalability by leveraging on a communication pattern that
does not require an all-to-all data exchange as our previous solution, whereas
the former improves the performance of each GPU. The combination of the two
enhancements provides a significant advantage with respect to our previous
solution: the number of Traversed Edges Per Second (TEPS) is four times greater
for large graphs that require at least 1024 GPUs.

As a further optimization, we adopt a technique to reduce the size of exchanged messages
that relies on the use of a bitmap (as in \cite{Satish:2012}). This change
halves, by itself, the total execution time.

The paper is organized as follows: in Section \ref{sec:background}, we discuss
our previous solution presented in \cite{Mastro2013} and \cite{Bernaschi2014}.
Our current work is presented in Section \ref{sec:multi_gpu}; Section
\ref{sec:results} reports the results obtained with up to 4096 GPUs.  In
section \ref{sec:compression} we describe the implementation of the bitmap to
reduce the communication data-size.  In Section \ref{sec:related_works} we
briefly review some of the techniques presented in related works. Finally
Section \ref{sec:conclusions} concludes the work with a perspective on future
activities.

\section{Background on parallel distributed BFS \label{sec:background}}

\subsection{Parallel distributed BFS on multi-GPU systems\label{distrib_bfs}}

In a distributed memory implementation, the graph is partitioned among the
computing nodes by assigning to each one a subset of the original vertices and
edges sets. The search is performed in parallel, starting from the processor
owning the root vertex. At each step, processors handling one or more frontier
vertices follow the edges connected to them to identify unvisited neighbors.
The reached vertices are then exchanged in order to notify their owners and a
new iteration begins. The search stops when the connected component containing
the root vertex has been completely visited.

The partitioning strategy used to distribute the graph is a crucial part of the
process because it determines the load balancing among computing nodes and
their communication pattern. Many authors reported that communications
represent the bottleneck of a distributed a BFS \cite{Yoo:2005, Buluc:2011,
Ueno:2012}.  For what concerns computations, using parallel architectures such
as GPUs, introduces a second level of parallelism that exploits a shared memory
approach for the local processing of the graph.

In our first work \cite{Mastro2013} we implemented a parallel distributed BFS
for multi-GPU systems based on a simple partitioning of the input graph where
vertices were assigned to processors by using a {\em modulo} rule. Such
partitioning resulted in a good load balancing among the processors but
required the processors to exchange data with, potentially, every other
processor in the pool. This aspect limited the scalability of the code beyond
$1024$ GPUs.

As to the local graph processing, in our original code we parallelized the
frontier expansion by using a GPU thread per edge connected to the frontier. In
that way each thread is in charge of only one edge and the whole next level
frontier set can be processed in parallel. In \cite{Mastro2013} we described a
technique to map threads to data that achieves a perfect load-balancing by
combining an {\em exclusive scan} operation and a {\em binary search} function.

The distributed implementation requires the data to be correctly arranged
before messages can be exchanged among computing nodes: vertices reached from
the frontier must be grouped by their owners. Moreover, many vertices are
usually reached from different frontier vertices \cite{Merrill:2012,
Beamer:2012, Mastro2013}, therefore it is important to remove duplicates before
data transfers in order to reduce communication overhead.

The removal of duplicates and the grouping of vertices can be implemented in a
straightforward way by using the {\em atomic} operations offered by CUDA.
However, at the time of our original work, atomics were quite penalizing so we
implemented the two operations by supporting benign race conditions using an
integer map and a parallel compact primitive \cite{Bernaschi2014}. The
scalability of the code, however, was still limited to 1024 nodes.

Different partitioning strategies can be employed to reduce the number of
communications \cite{Yoo:2005, Buluc:2011, Satish:2012}. Hereafter, we present
the results obtained by combining a 2D partitioning scheme of the adjacency
matrix representing the graph with our frontier expansion approach that takes
advantage of the improved performance of the atomic operations available with
the Nvidia Kepler architecture.

\subsection{2D Partitioning}\label{2DPart}

\rem{In the 1D partitioning scheme implemented in our original code (Section
\ref{sec:background}) the processors handle the whole adjacency list for each
of their vertices.  For this reason the frontier expansion step can lead to any
other processor in the pool thus requiring an all-to-all communication pattern
that limits the overall performances.

In the present work, we report preliminary results obtained by combining a 2D
decomposition strategy for the adjacency matrix with the mapping between data
and threads peculiar to our previous code.}

\begin{figure}[t!]
\begin{center}
\includegraphics[scale=0.3]{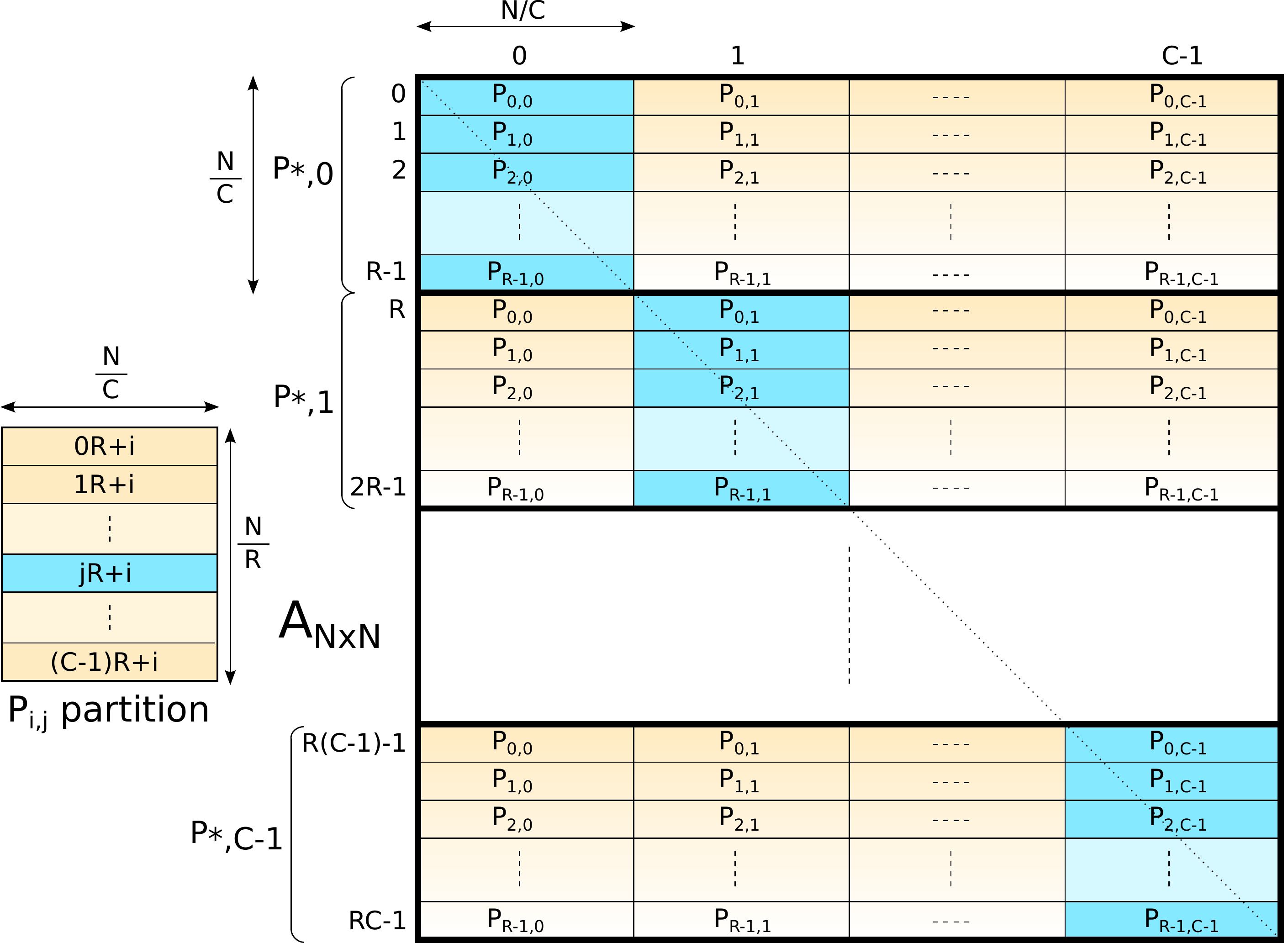}
\end{center}
\caption{Two-dimensional partitioning of an adjacency matrix $A$ with an
$R\times C$ processor grid. The matrix is divided into $C$ consecutive groups
of $R\times C$ blocks of edges along the vertical direction. Each block is a
$N/(RC) \times N/C$ sub-matrix of $A$. Different groups of blocks are colored
with different yellow gradients. For each block, the column of processors
owning the corresponding vertices (row indexes) are shown in blue. On the left
part, it is shown the sequence of blocks, from top to bottom, assigned to the
generic processor $(p_i,p_j)$. The colors correspond to the blocks assigned to
the processor in each group of blocks.}
\label{fig:2D_dec}
\end{figure}

We implemented a 2D partitioning scheme similar to that described by Yoo {\em
et al.} in \cite{Yoo:2005}, where $RC$ computing nodes are arranged as a
logical grid with $R$ rows and $C$ columns and mapped onto the adjacency matrix
$A_{N\times N}$, partitioning it into blocks of edges. The processor grid is
mapped once horizontally and $C$ times vertically thus dividing the columns in
$C$ blocks and the rows in $RC$ blocks, as shown in Figure \ref{fig:2D_dec}.

\noindent Processor $P_{ij}$ handles all the edges in the blocks $(mR+i,~j)$,
with $m=0,...,C-1$. Vertices are divided into $RC$ blocks and processor
$P_{ij}$ handles the block $jR+i$. Considering the edge lists represented along
the columns of the adjacency matrix, this partitioning is such that:

\begin{itemize}
\item[{\em (i)}] the edge lists of the vertices handled by each processor are
partitioned among the processors in the same grid column;
\item[{\em (ii)}] for each edge, the processor in charge of the destination
vertex is in the same grid row of the edge owner.
\end{itemize}

\begin{algorithm}[ht] 
\caption{\label{alg:bfs2d} Parallel BFS with 2D partitioning.} 
\begin{algorithmic}[1]
\scriptsize
\REQUIRE Root vertex $r$.
\REQUIRE Processor $P_{ij}$.
\\~ 

\STATE level[:] $\gets -1$
\STATE pred[:] $\gets -1$
\STATE bmap[:] $\gets 0$
\STATE front $\gets \emptyset$
\IF {r belongs to processor $P_{ij}$}
        \STATE level[r] $\gets 0$
        \STATE pred[r] $\gets r$
        \STATE bmap[r] $\gets 1$
        \STATE front = $\left\{r\right\}$
\ENDIF
\STATE lvl $\gets 1$
\WHILE {true}
        \STATE front $\gets$ {\bf gather} front[] from column $j$ //vertical comm
        \FOR {{\bf each} u {\bf in} front}
                \FOR {{\bf each} (u,v) {\bf in} local edge list}
                        \STATE col $\gets$ column of v's owner
                        \STATE {\bf send} (u,v) to processor $P_{i,col}$ //horizontal comm
                \ENDFOR
        \ENDFOR
        \STATE front $\gets \emptyset$
        \FOR {{\bf each} received edge (u,v)}
                \IF {bmap[v] = 0}
                        \STATE bmap[v] $\gets 1$
                        \STATE pred[v] $\gets$ u
                        \STATE level[v] $\gets$ lvl
                        \STATE front = front $\cup$ \{v\}
                \ENDIF
        \ENDFOR
        \STATE lvl $\gets$ lvl+1
        \IF {front = $\emptyset$ {\bf for all} processors}
                \STATE {\bf break}
        \ENDIF
\ENDWHILE
\end{algorithmic}
\end{algorithm}

With a similar decomposition, each step of the BFS requires two communication
phases, called {\em expand} and {\em fold}. The first one involves the
processors in the same grid column whereas the second those in the same grid
row.  Algorithm \ref{alg:bfs2d} shows a pseudo code for a parallel BFS with 2D
decomposition.  At the beginning of the each step, each processor has its own
subset of the frontier set of vertices (initially only the root vertex). The
search entails the scanning of the edge lists of all the frontier vertices. Due
to property {\em (i)} of the 2D decomposition, each processor gathers the
frontier sets of vertices from the other processors in the same
processor-column (vertical exchange, line 13).  The frontier is then expanded
by having each column of processors to collectively scan the edge lists of the
gathered sets of vertices in search of edges leading to unvisited neighbors.
For property {\em (ii)}, edges found are sent to the processors, in the same
grid row, that own the destination vertices (horizontal exchange, lines 14-19).
Unvisited destination vertices of the received edges form the frontier of the
next BFS step (lines 20-28). The search ends when the frontier of each
processor is empty, meaning that the whole connected component containing the
root vertex has been visited \cite{2DAlgOpt}.

The main advantage of the 2D partitioning is a reduction of the number of
communications. If $P$ is the number of processors, our first implementation
required $O(P)$ data transfers at each step whereas the 2D partitioning only
requires $2\times O(\sqrt{P})$ communications.

\section{BFS on a multi-GPU system with 2D partitioning\label{sec:multi_gpu}}

Our work loosely follows the Graph500 \cite{graph500} benchmark specifications.
The benchmark requires to generate in advance a list of edges with an R-MAT
generator \cite{RMAT} and to measure the performances over 64 BFS operations
started from random vertices. It poses no constraint about the kind of data
structures used to represent the graph but it requires the vertices to be
represented using at least 48 bits. Since the focus of the present work is the
evaluation of the new local part of the search and the 2D partitioning with
respect to our original code, we do not strictly adhere to the Graph500
specifications and represent vertices with 32 bits\footnote{This choice does
not limit the total size of graphs. They are generated or read by using 64 bits
per vertex. The 32-bit representation is used to store local partitions.} (more
than enough to index the highest number of vertices storable in the memory of
current GPUs).

\subsection{Local graph data structure\label{DataStructs}}

Each processor stores its part of the adjacency matrix as a $(N/R)\times (N/C)$
local matrix (Figure \ref{fig:2D_dec}) where blocks of edges are stored in the
same row order as in the global matrix. This allows to map global indexes to
local indexes so that global row $v$ is mapped to the same local row for every
processor in the same processor-row of the owner of vertex $v$.  In a similar
way, global column $u$ is mapped to the same local column for every processor
in the same processor-column handling the adjacency list of vertex $u$.

Local adjacency matrices are sparse as the global matrix, so they are stored in
compressed form. Since they are accessed by reading an entire column for each
vertex of the frontier, we used a representation that is efficient for column
access, the Compressed Sparse Column (CSC) format.  As a consequence, processors
may scan adjacency lists by accessing blocks of consecutive memory locations
during the frontier expansion (one block for each vertex in the frontier of the
same processor-column). Since the non-zeroes entries of an adjacency matrix have
all the same value, the CSC is represented with only two arrays: the column
offset array \verb|col| and the row index array \verb|row|.

Predecessors and BFS levels are stored in arrays of size $N/R$ (number of rows
of the local matrix). For the frontier, since it can only contain local
vertices, we use an array of size $N/(RC)$ and the information about visited
vertices is stored in a bitmap with $\lceil \frac{N/R}{32} \rceil$ 32-bit words.

\subsection{Parallel multi-GPU implementation\label{subsec:2Dalgo}}

The code may generate a graph by using the R-MAT generator provided by the
reference code available from the Graph500
website\footnote{http://www.graph500.org/referencecode}. Then the graph is
partitioned as described in Section \ref{2DPart}. The BFS phase is performed
entirely by the GPUs with the CPUs acting as network coprocessors to assist in
data transfers.

Algorithm {\ref{alg:bfs2dDet}} shows the code scheme of the BFS phase.  Every
processor starts with the level, predecessor and bitmap arrays set to default
values (lines 1-4). The owner of the root vertex copies its local column index
into the frontier array and sets its level, predecessor and visited bit (lines
5-10). All the data structures are indexed by using local indexes.  At the
beginning of each step of the visit, the {\bf expand} communication is
performed (line 13). Every processor exchanges its frontier with the other
processors in the same processor-column and stores the received vertices in the
\verb|all_front| array.  Note that in this phase, since processors send subsets
of their own vertices, only disjoint sets are exchanged.  After the exchange,
in the {\bf expand\_frontier} routine, processors in each column collectively
scan the whole adjacency lists of the vertices in \verb|all_front| (line 14).
For each vertex, its unvisited neighbors are set as visited and the vertex is
set as their predecessor. For neighbors owned locally the level is also set.
The routine returns the unvisited neighbors, divided by the processor-column of
the owner, inside an array with $C$ blocks, \verb|dst_verts|.

\begin{algorithm}[H]
\caption{\label{alg:bfs2dDet} GPU version of Parallel BFS.}
\begin{algorithmic}[1]
\footnotesize
\REQUIRE Root vertex $r$.
\REQUIRE CSC=(row[],col[]).
\REQUIRE Processor $P_{ij}$.
\\~

\STATE level[:] $\gets -1$
\STATE pred[:] $\gets -1$
\STATE bmap[:] $\gets 0$
\STATE front $\gets []$
\IF {r belongs to $P_{ij}$}
        \STATE level[LOCAL\_ROW(r)] $\gets 0$
        \STATE pred[LOCAL\_ROW(r)] $\gets r$
        \STATE bmap[LOCAL\_ROW(r)] $\gets 1$
        \STATE front[0] $\gets$ LOCAL\_COL($r$)
\ENDIF
\STATE lvl $\gets 1$
\WHILE {true}
        \STATE all\_front $\gets$ {\bf expand\_comm}(front)
        \STATE dst\_verts $\gets$ {\bf expand\_frontier}(row, col, level, pred, bmap, all\_front)
        \STATE front[:] $\gets$ dst\_verts[$j$][:]
        \STATE dst\_verts[$j$] $\gets$ []
        \STATE int\_verts $\gets$ {\bf fold\_comm}(dst\_verts)
        \STATE front $\gets$ front $\oplus$ {\bf update\_frontier}(row, col, level, pred, bmap, int\_verts)
        \IF {len(front) = $0$ for all processors}
                \STATE {\bf break}
        \ENDIF
        \STATE lvl $\gets$ lvl$+1$
\ENDWHILE
\end{algorithmic}
\end{algorithm}

\noindent
After the frontier expansion, neighbors local to the processor are moved from
the $j$-th block of the \verb|dst_verts| array into the \verb|front| array
(lines 15-16) and the {\bf fold} communication phase is performed. Unvisited
neighbors just discovered are sent to their owners, located in the same
processor-row, and received vertices are returned in the \verb|int_verts| array
(line 17).  Finally, the {\bf update\_frontier} routine selects, among the
received vertices, those that have not been visited yet and sets their level and
bitmap bit (line 18). Returned vertices are then appended to the new frontier
and the cycle exit condition is evaluated (line 19). The BFS continues as long
as at least one processor has a non-empty frontier at the end of the cycle.

The output is validated by using the same procedure included in our original
code.

\subsection{Communications}

The expand and fold communication phases are implemented with point-to-point
MPI primitives and make use of the following scheme:

\begin{itemize}
\item start {\em send} operations;
\item {\em wait} for completion of receive operations posted in the previous round;
\item post non-blocking {\em receive} operations for the next round.
\end{itemize}

\noindent
that hides the latency of the receive operations in the BFS round and
avoids possible deadlocks due to the lack of receive buffers.

Since the communication involves only local indexes, the expand and fold
phases are carried out by exchanging 32-bit words.

\subsection{Frontier expansion}

After the local frontiers have been gathered, the frontier expansion phase
starts. This phase has a workload proportional to the sum of the degrees of the
vertices in the current frontier. There are no floating point operations, just
few integer arithmetic operations, mainly for array indexing, and it is memory
bandwidth bound with an irregular memory access pattern. It is characterized by
a high degree of intrinsic parallelism, as each edge originating from the
frontier can be processed independently from the others.  There could be,
however, groups of edges leading to the same vertex. In those cases, only one
edge per group should be processed. In our code we use a thread for each edge
connected to the frontier and this allows the selection of the single edge to be
followed in, at least, two ways: either by taking advantage of benign race
conditions or by using synchronization mechanisms such as atomic operations.

In our previous work, we chose to avoid atomic operations because they were
quite penalizing with the Nvidia Fermi architecture, available at that time.
The current architecture, code-named Kepler, introduced many enhancements
including a significant performance improvement of the atomic operations (almost
an order of magnitude with respect to Fermi GPUs). As a consequence, we decided
to use atomics in the new 2D code.  The resulting code is much simpler because
there is no longer the need to support the benign race conditions on which the
original code relied \cite{Mastro2013}. Atomics are used to access the bitmap in
both the frontier expansion and update phases and to group outgoing vertices
based on destination processors.

\begin{figure}[t]
\begin{center}
\includegraphics[scale=0.5]{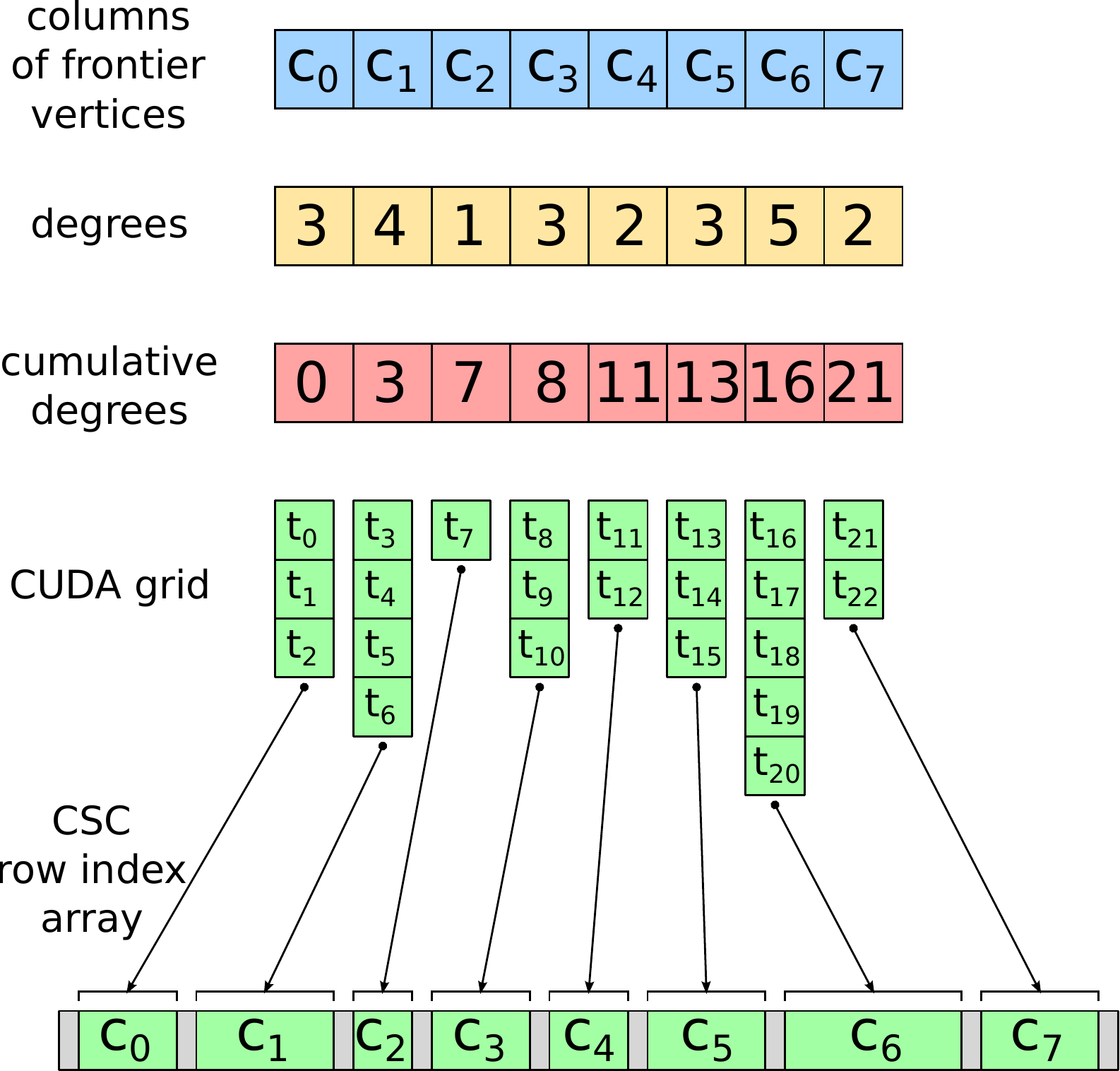}
\end{center}
\caption{Threads to data mapping assigning one thread per edge. From top to
bottom, the first array represents the local column indexes corresponding to
frontier vertices. The second contains the local degrees of each vertex, {\em
i.e.} the number of non-zeroes per column (23 in total), and the third one is
the cumulative degree array. The CUDA grid is launched with (at least) 23
threads.  By searching the cumulative array for the greatest entry less than its
global id, each thread finds its frontier vertex. Finally, threads mapped to the
same frontier vertex process its edge list (a CSC column).} \label{fig:t2d_map}
\end{figure}

At the beginning of the expansion phase, the vertices in the blocks of the
\verb|all_front| array are copied in a contiguous block of device memory and
then the corresponding columns of the CSC matrix are processed. We employ a
mapping between data and threads similar to the one used in our previous code
\cite{Mastro2013} where a CUDA thread is used for every edge originating from
the current frontier. The mapping allows to achieve an optimal load balancing as
edge data are evenly distributed among the GPU threads. Given the frontier
vertices $u_0,u_1,u_2,...,u_{n-1}$ with degrees $d_0,d_1,d_2,...,d_{n-1}$ and
adjacency lists:

\[
v^0_0,v^0_1,...,v^0_{d_0-1}|v^1_0,v^1_1,...,v^1_{d_1-1}|...|v^{n-1}_0,v^{n-1}_1,
...,v^{n-1}_{d_{n-1}-1}
\]

\noindent
thread $i$ is mapped onto the $j$-th edge connected to vertex $u_k$:

\[
i~~\leftrightarrow~~(u_k,v^k_j)
\]
\[k=max\left\{l~\middle|~\sum_{s=0}^{l} d_s
\le i,~\forall l < n \right\}~~~~j=i-\sum_{s=0}^{k} d_s
\]

\noindent
After vertices in \verb|all_front| are copied to device memory, we use their
degrees to compute a cumulative degree array by means of an exclusive scan
operation.
The frontier expansion kernel is then launched with $r=\sum_{s=0}^{n-1} d_s$ threads.
Each thread processes one local edge $(u,v)$ originating from the frontier, {\em
i.e.} one element of column $u$ of the CSC (see Figure \ref{fig:t2d_map}).
Algorithm \ref{alg:front-exp} shows the pseudo code for the frontier expansion kernel.
The source vertex $u$ is found by performing a binary search for the greatest
index $k$ such that \verb|cumul[k]| is less than or equal to the global thread id
(line 2) and by accessing the $k$-th location of the \verb|frontier| array
(line 3) \cite{BinSearchAcc}.

The column offset of vertex $u$ is found in the $u$-th element of the CSC column
index array \verb|col|. The row offset of the destination vertex is computed by
subtracting from the thread id the degrees of the vertices preceding $u$ in the
\verb|frontier| array. Finally, the local id $v$ of the destination vertex is
found by accessing the CSC row index array at the location corresponding to the
sum of both the column and the row offsets (line 4).

The status of the neighbor $v$ is then checked (lines 5-6).  If the vertex has
already been visited, then the thread returns. Otherwise, it is marked as
visited with an {\em atomicOr} operation (line 7). That instruction returns the
value of the input word before the assignment. By checking the vertex bit in the
return value, it is possible to identify the first of different threads handling
edges leading to the same vertex (assigned to different columns) that set the
visited bit (line 8). Only the first thread proceeds, the others return.

The bitmap is used not only for the vertices owned locally but also for those,
belonging to other processors (in the same processor-row), that are reachable
from local edges. In other words, the bitmap has an entry for each row of the
CSC matrix. This makes possible to send external vertices just once in the fold
exchanges because those vertices are sent (and marked in the bitmap) only when
they are reached for the first time in the frontier expansion phase.

\begin{algorithm}[H]
\caption{\label{alg:front-exp} CUDA atomic-based frontier
expansion kernel.}
\begin{algorithmic}[1]
\footnotesize
\REQUIRE frontier[].
\REQUIRE current level.
\REQUIRE dst\_cnt[]=0.
\REQUIRE cumulative degree array cumul[].
\REQUIRE Processor $P_{ij}$.
\REQUIRE CSC=(row[],col[]), bmap[], level[] and pred[].
\\~

\STATE gid $\gets$ (blockDim.x*blockIdx.x + threadIdx.x)
\STATE k $\gets$ binsearch\_maxle(cumul, gid)
\STATE u $\gets$ frontier[k]
\STATE v $\gets$ row[col[u] + gid - cumul[k]]
\STATE m $\gets$ 1 \verb|<<| (v $\bmod$ 32)
\STATE {\bf if} (bmap[v/32] \& m) {\bf return}
\STATE q $\gets$ atomicOr(\&bmap[v/32], m)
\IF {!(m \& q)}
        \STATE $tgtj \gets$ v/(N/(RC))
        \STATE off $\gets$ atomicInc(\&dst\_cnt[$tgtj$])
        \IF {($tgtj$ != $j$)}
                \STATE dst\_verts[$tgtj$][off] $\gets$ v
        \ELSE
                \STATE dst\_verts[$j$][off] $\gets$ ROW2COL(v)
                \STATE lvl[v] $\gets$ level
        \ENDIF
        \STATE pred[v] $\gets$ u
\ENDIF
\end{algorithmic}
\end{algorithm}

\noindent
Each unvisited neighbor is added to the array associated to the
processor-column of its owner processor in preparation for the subsequent fold
exchange phase (lines 12 and 14). The processor-column is computed based on the
local index of the neighbor (line 9) and the counter for such column is
atomically incremented to account for multiple threads appending for the same
processor (line 10). If the neighbor belongs to the local processor its level
is also set (line 15).  Finally, regardless of the owner, the neighbor
predecessor is set (line 17).

After the kernel has completed, the array containing the discovered vertices
(grouped by owner) is copied to host memory in preparation for the fold
communication phase.

The bitmap, level and predecessor arrays are never copied back to the host
because they are used only by GPU kernels.

\subsubsection{Optimizations}

In the current CUDA implementation, we introduced an optimization to mitigate
the cost of the binary searches performed at the beginning of the frontier
expansion kernel.
Since threads search for their global index in the cumulative array,
non-decreasing indexes are returned to consecutive threads:

\[
\verb|binsearch_maxle(cumul, gid)| \le
\]
\[
\verb|binsearch_maxle(cumul, gid+1)|
\]
Where \verb|gid| is the global thread identifier that is equal to:
\[
\verb|threadIdx.x + blockIdx.x * blockDim.x|
\]

That relation makes possible to scan the columns with fewer threads
than the number of edges originating from the frontier without
increasing the number of binary searches performed {\em per} thread.
This is done by assigning each thread to a group of consecutive
elements in the CSC columns. If not enough elements are available in
the column, the group overlaps on the next column.  Then, each thread
performs a binary search only for its first edge to obtain a base
column index. The indexes for subsequent edges $tid+i$ are found by
incrementing the base index until $(tid+i) \ge cumul[base+1]$.  The
overhead introduced by these linear searches is very limited because
the majority of edge groups are contained in a single column and so
the base column index is never incremented.

Among other optimizations, there is the passage of read-only arrays to kernels
with the modifiers {\em const/restrict}. In this way the compiler uses the
read-only data cache load functions to access them.  Moreover, in order to
increase instruction-level parallelism, edges assigned to a thread are not
processed sequentially, one after the other. Instead, they are processed
together in an interleaved way by replacing each kernel instruction with a
sequence of independent instructions, one for each edge. In this way, since
threads do not stall on independent instructions, each group is executed in a
way that hides arithmetic and memory latencies.

\begin{figure}
\begin{center}
\includegraphics[scale=0.5]{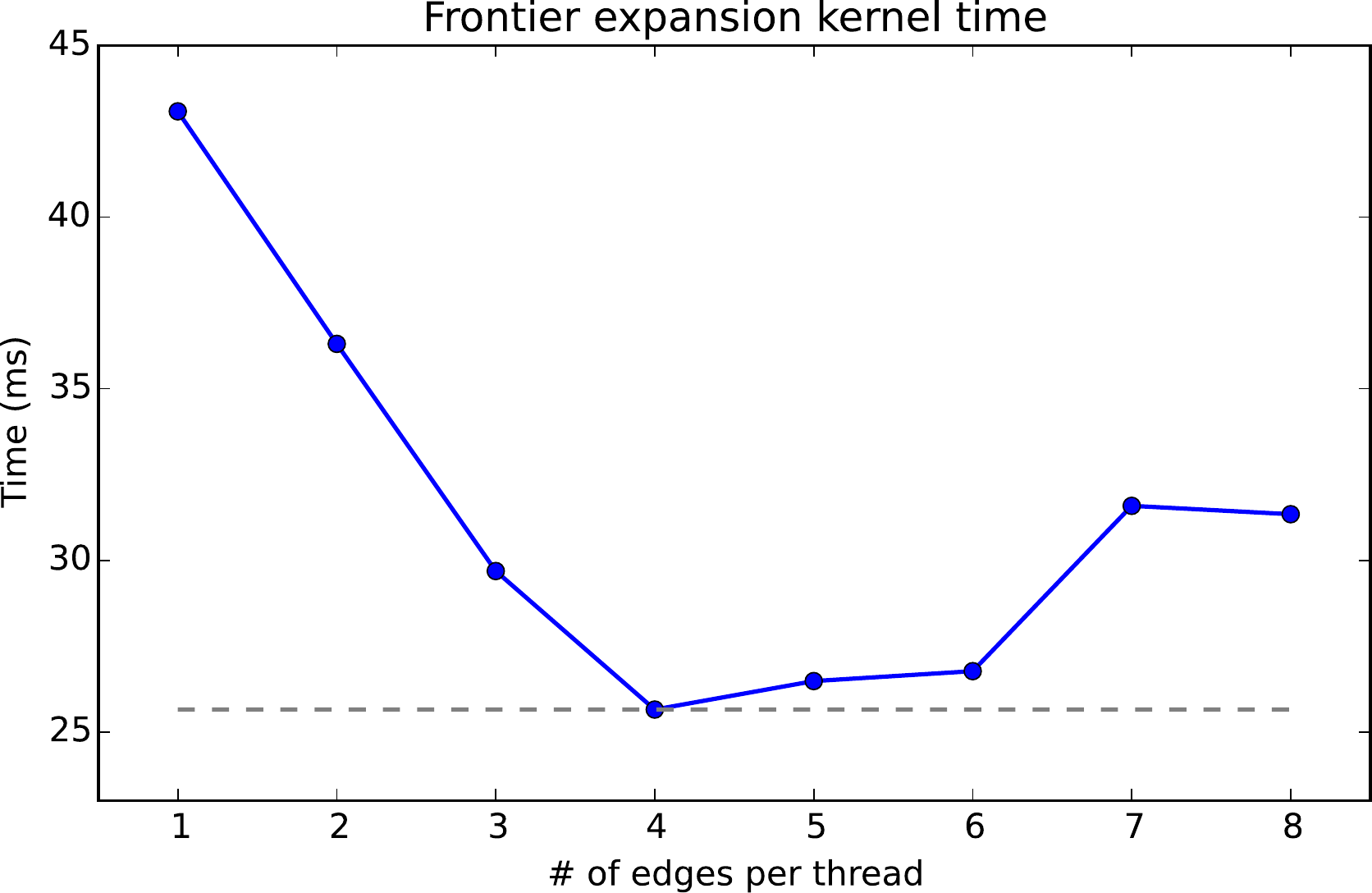}
\end{center}
\caption{Frontier expansion times measured visiting an R-MAT graph with
scale=${21}$ using a single GPU and varying the number of edges per GPU
thread.}\label{fig:multi_edge}
\end{figure}

Figure \ref{fig:multi_edge} reports the total time spent in the frontier
expansion kernel to perform single-GPU BFSs with a variable number of edges
per thread on an R-MAT graph with scale=$21$ . The maximum performance is obtained by
assigning four edges per thread with a performance gain of ${\sim}40\%$
with respect to the case with a thread per neighbor.

\subsection{Frontier update}

In the frontier update phase, local vertices remotely discovered and received
during the fold exchange, are processed to find those that are locally
unvisited, {\em i.e.}, whose visited bit is not set.

Since this phase processes the vertices received, instead of their adjacency
lists, fewer computational resources are required compared to those necessary
in the frontier expansion phase.

As in the expand communication phase, vertices returned by the fold exchange
are grouped by sender in different blocks and thus, before starting the
processing, they are copied to device memory in contiguous locations.

After the copy, the frontier update kernel is launched with a thread per vertex.
Threads are mapped linearly onto the arrays of vertices and each thread is
responsible of updating the level and predecessor of its vertex, if unvisited,
and to add it to the output array. As in the frontier expansion kernel, we make
use of atomic operations in order to synchronize the accesses to the bitmap and
the writes to the output buffer.
As to the predecessor, since in the fold phase we do not transmit source
vertices, the thread, lacking this information, stores the sender
processor-column in the predecessor array (the sender saved the predecessor).

After the kernel has completed, the output array is copied to host memory and it
is appended to the frontier array.

\section{Results \label{sec:results}}

The performances hereafter reported have been obtained on the Piz Daint system
belonging to the Centro Svizzero di Calcolo Scientifico (CSCS). Piz Daint is a
hybrid Cray XC30 system with 5272 computing nodes interconnected by an Aries
network with Dragonfly topology. Each node is powered by both an Intel Xeon
E5-2670 CPU and a NVIDIA Tesla K20X GPU and is equipped with 32 GB of DDR3 host
memory and 6 GB of GDDR5 device (GPU) memory.
The code has been built with the GNU C compiler version 4.8.2, CUDA C compiler
version 5.5 and Cray MPICH version 6.2.2. For the GPU implementation of the
exclusive scan we used the CUDA Thrust library \cite{THRUST}, a well-known, high
performance template library implementing a rich collection of data parallel
primitives.

\begin{figure}[H]
\begin{center}
\includegraphics[scale=0.5]{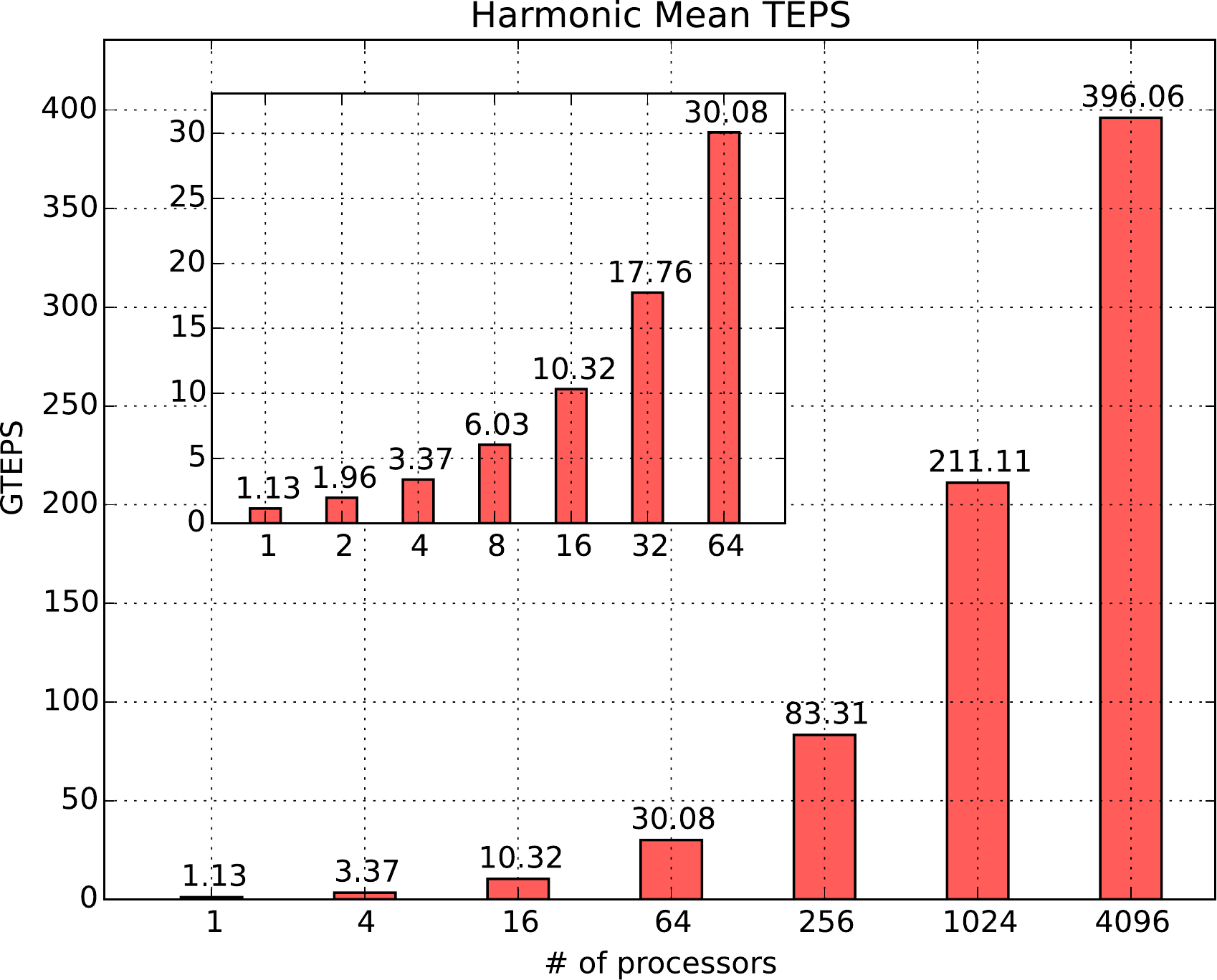}
\end{center}
\caption{Harmonic mean TEPS measured with a number of GPUs ranging from 1 to
4096 keeping the graph scale per processor fixed. \label{fig:mean_teps}}
\end{figure}

\begin{figure}[H]
\begin{center}
\includegraphics[scale=0.5]{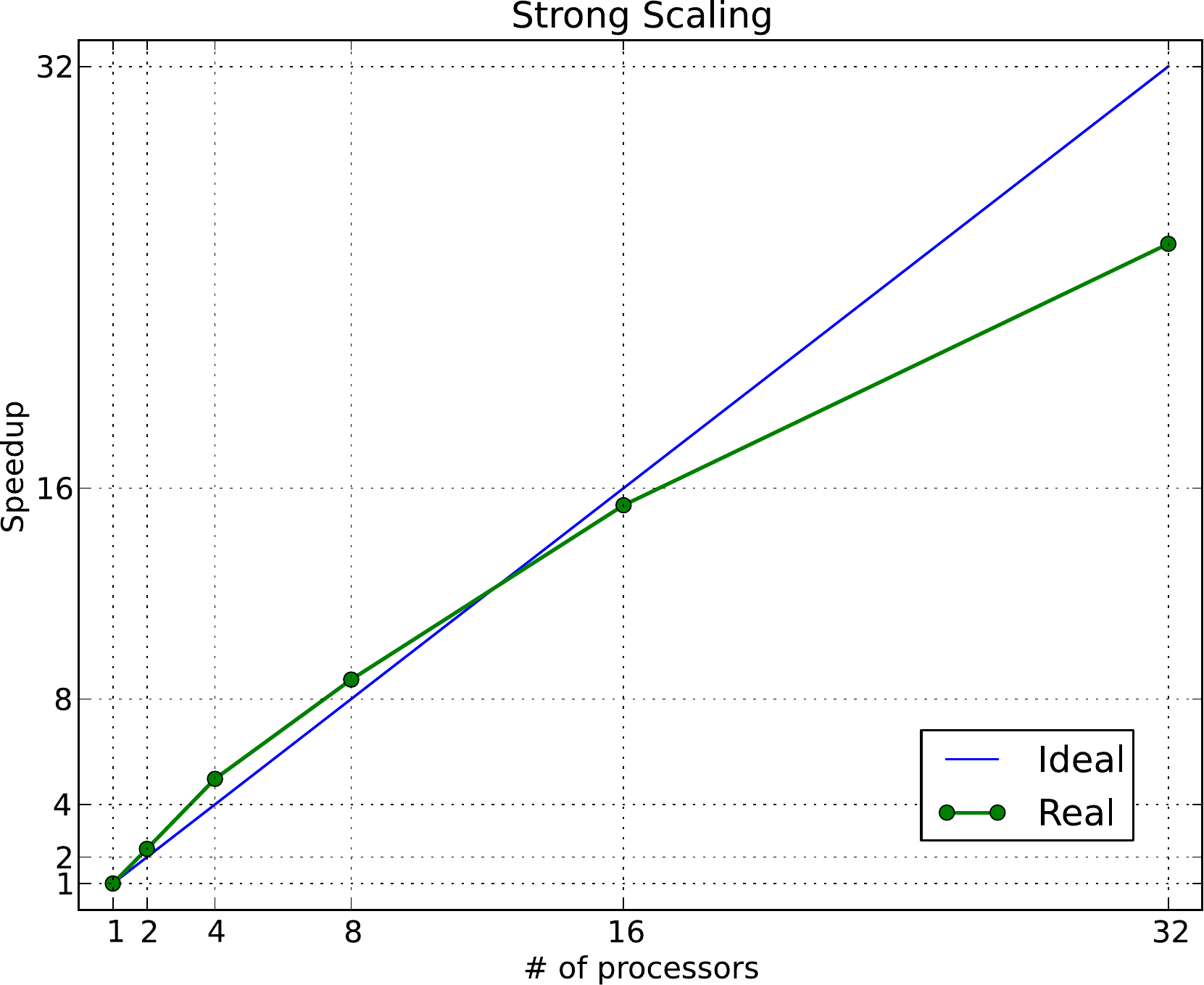}
\end{center}
\caption{Speedup of the 2D code measured by visiting an R-MAT graph with
scale=${25}$.}\label{fig:strong_sc}
\end{figure}

\begin{table}[H]
\begin{center}
\centering
\resizebox{0.8\textwidth}{!}{
\begin{tabular}{|c|c|c|c|||c|c|c|c|}
        \hline
        \# of GPUs & grid size & scale & edge factor & \# of GPUs & grid size & scale & edge factor\\\hline
        1 & $1\times 1$ & 21 &    &   128 & $ 8\times 16$ & 28 &   \\
        2 & $1\times 2$ & 22 &    &   256 & $16\times 16$ & 29 &   \\
        4 & $2\times 2$ & 23 &    &   512 & $16\times 32$ & 30 &   \\
        8 & $2\times 4$ & 24 & 16 &  1024 & $32\times 32$ & 31 & 16\\
       16 & $4\times 4$ & 25 &    &  2048 & $32\times 64$ & 32 &   \\
       32 & $4\times 8$ & 26 &    &  4096 & $64\times 64$ & 33 &   \\
       64 & $8\times 8$ & 27 &    &       &               &    &   \\\hline
\end{tabular}
}
\end{center}
\caption{Processor grid and R-MAT graph configurations used for the tests.}
\label{tab:runconf}
\end{table}

The code uses 32-bit data structures to represent the graph because the memory
available on a single GPU is not sufficient to hold $2^{32}$ or more edges and
it transfers 32-bit words during the communications because the frontier
expansion/update and expand/fold communications work exclusively on local
vertices. We use 64-bit data only for graph generation/read and partitioning.
For the generation of R-MAT graphs we use the {\em make\_graph} routine found in
the reference code for the Graph500 benchmark. The routine returns a directed
graph with $2^{scale}$ vertices and ${\sim}edge\_factor\times 2^{scale}$ edges.
We turn the graph undirected by adding, for each edge, its opposite. Most of the
following results are expressed in Traversed Edges Per Second (TEPS), a
performance metric defined by the Graph500 group as the number of input edge
tuples within the component traversed by the search, divided by the time
required for the BFS to complete, starting right after the graph partitioning.
The tests have been performed with both R-MAT generated and real-world graphs.
Table \ref{tab:runconf} reports the processor-grid configurations, scales and
edge factors used for R-MAT graphs.

Figure \ref{fig:mean_teps} shows the weak scaling plot obtained by using a
number of GPUs ranging from 1 to 4096. In order to maintain  a fixed problem
size {\em per} GPU, the graph scale has been increased by 1 for each doubling
of the number of processors, starting from scale 21 whereas the edge factor had
been set equal to 16 (we set the scale equal to 21 because it is the maximum
value that can be used in the 1D implementation, against which we compare many
results).  For each scale, we report the harmonic mean of the TEPS measured in
64 consecutive BFS operations (a different root vertex is chosen at random for
each search). The code scales up to 4096 GPUs where the performances reaches
${\sim}400$ GTEPS on a graph with $2^{33}$ vertices and ${\sim}280$ billions of
directed edges. Figure \ref{fig:strong_sc} shows the strong scaling plot
obtained by visiting a fixed R-MAT graph with scale 25 and edge factor 16 with
an increasing number of GPUs. Here we used the maximum scale that
fits on a single GPU, to have more work to be done when the number
of processes increases.
The code scales linearly up to 16 GPUs. With 32 processors, the cost of data
transfers becomes comparable to the cost of computations and with more GPUs the
advantage becomes marginal and the efficiency drops.

\begin{figure}[H]
\begin{center}
\includegraphics[scale=0.4]{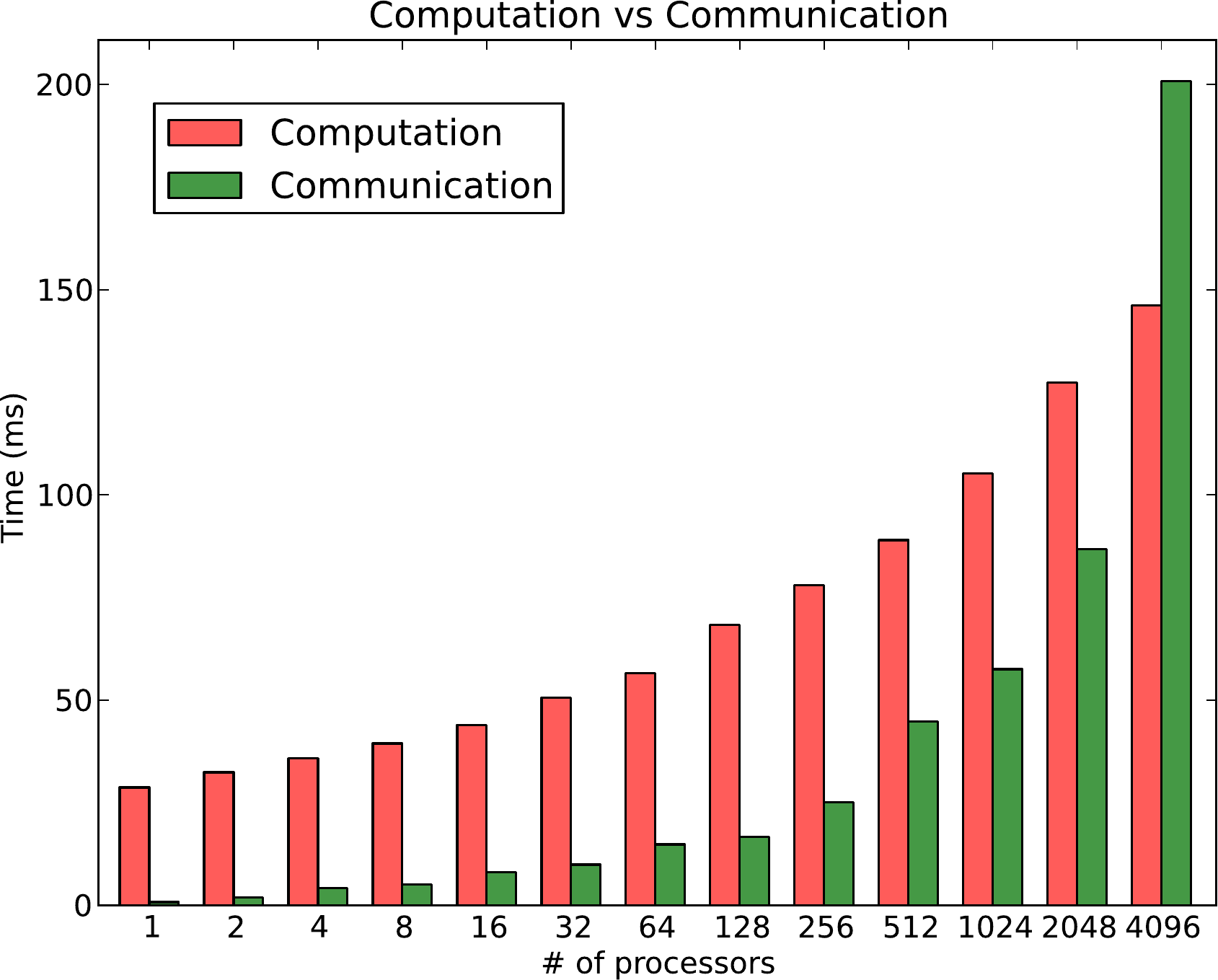}
\end{center}
\caption{Computation and transfer times (left and right bar, respectively) for
BFS operations performed with variable number GPUs on R-MAT graphs.}
\label{fig:comp_comm}
\end{figure}

\begin{figure}[H]
\begin{center}
\includegraphics[scale=0.4]{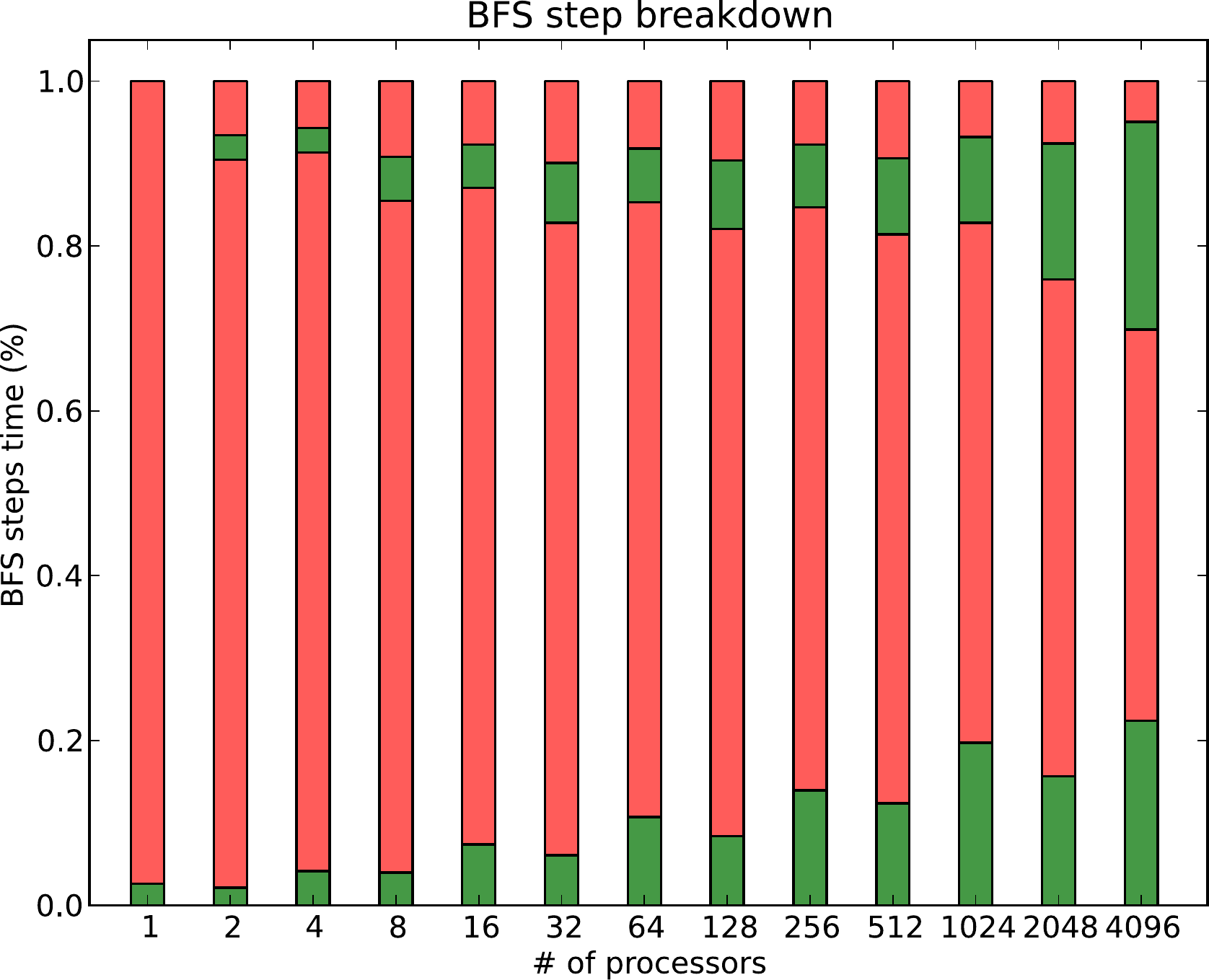}
\end{center}
\caption{Breakdown plot of the four phases of the BFS (from the bottom to the
top of the bars): expand exchange, frontier expansion, fold exchange and
frontier update. \rem{this plot does not report time spent in the allreduce at
the end of the BFS loop.}}\label{fig:breakdown}
\end{figure}

\begin{figure}[H]
\begin{center}
\includegraphics[scale=0.4]{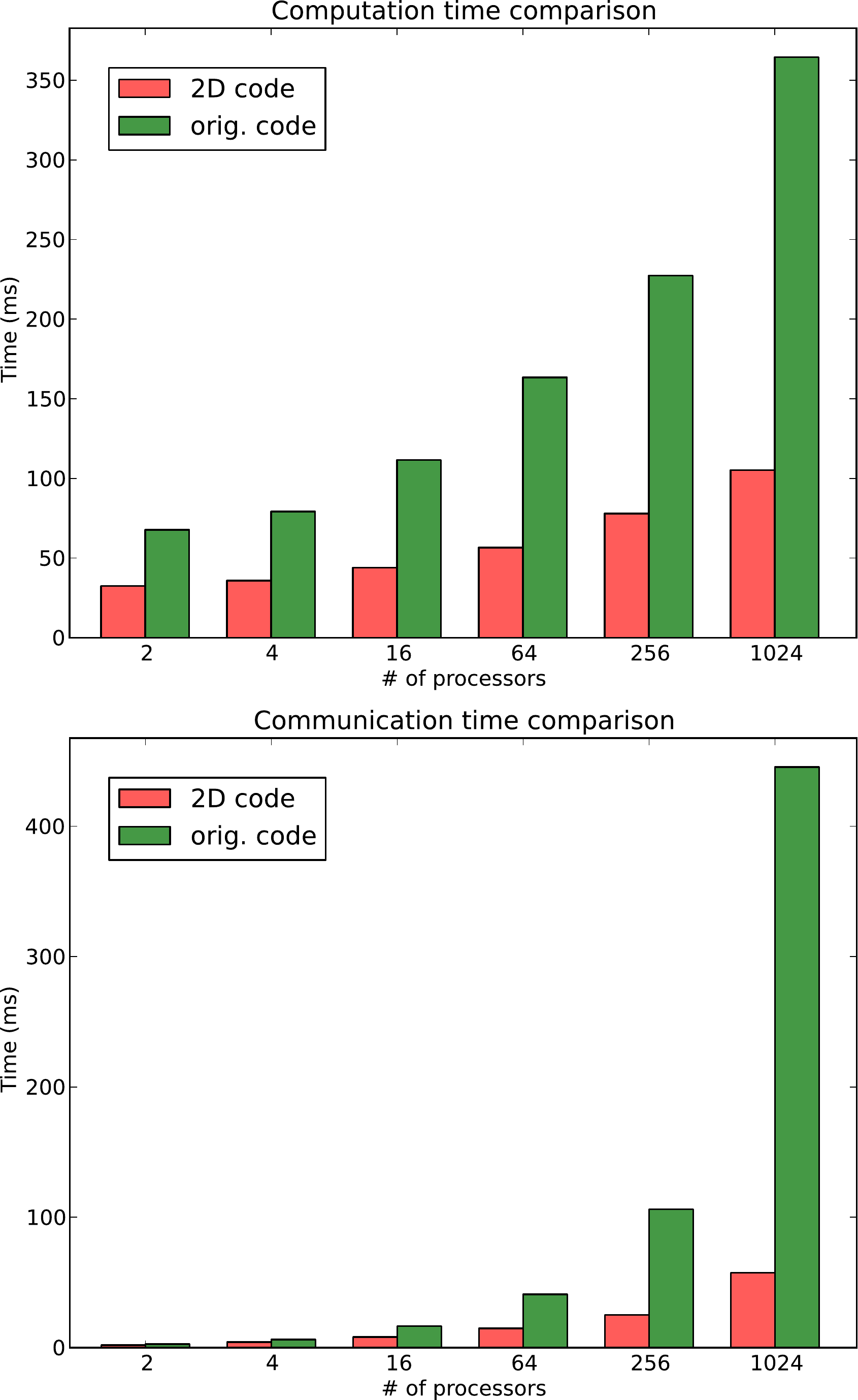}
\end{center}
\caption{Comparison of computation (upper plot) and communication (lower plot)
times between the new code based on 2D decomposition and our original code.
Both codes have been run on graphs as specified in the Table \ref{tab:runconf}.}
\label{fig:1D2D_comp}
\end{figure}

\begin{figure}[H]
\begin{center}
\includegraphics[scale=0.4]{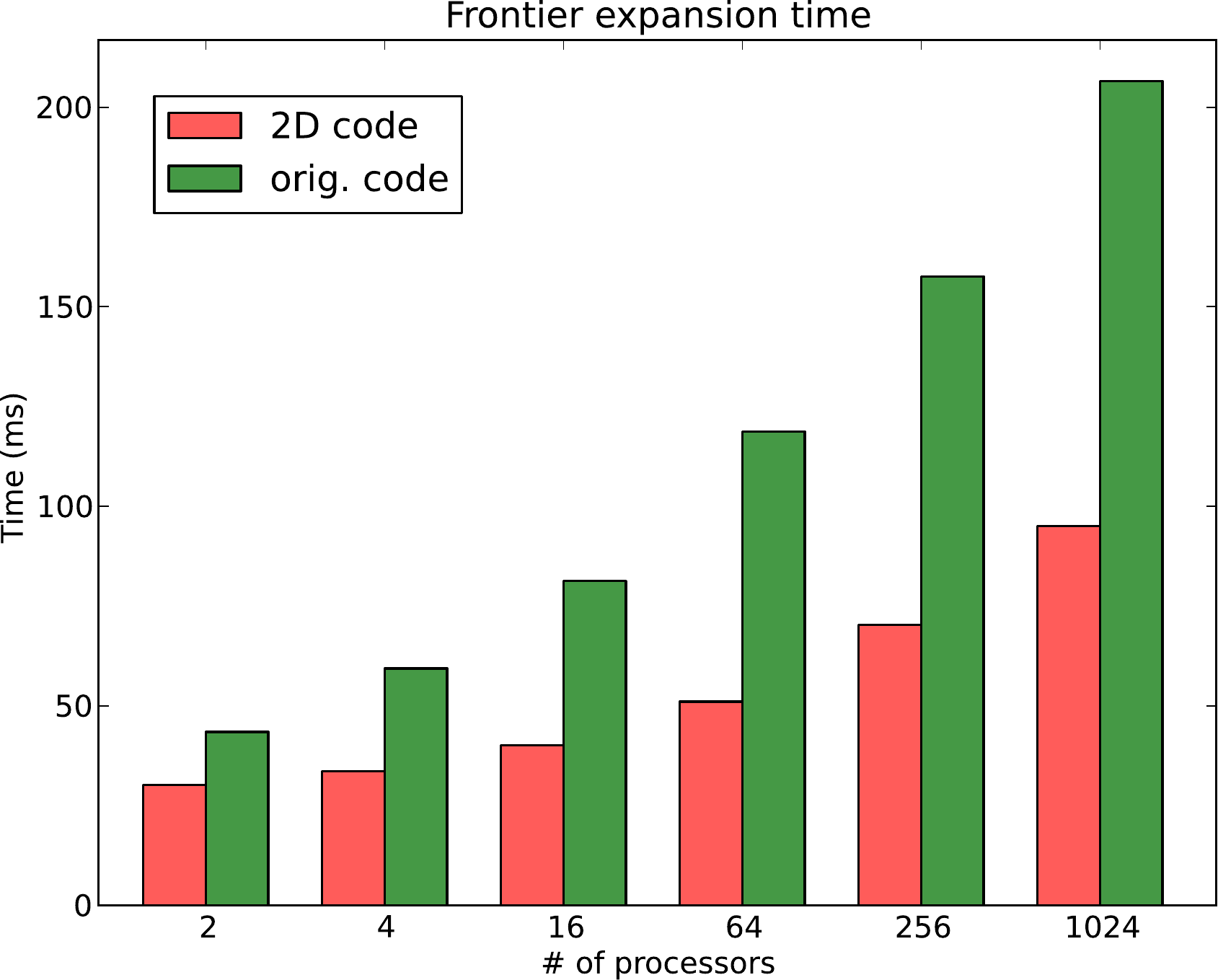}
\end{center}
\caption{Comparison of the time required by the frontier expansion phase in the
new code and the original code. The former is based on atomic operations whereas
the latter uses scatter/compact parallel primitives to support benign race
conditions.}\label{fig:1D2D_front_exp}
\end{figure}

\begin{table}[ht]
\begin{center}
\begin{tabular}{|c|c|c|}
        \hline
        \# of GPUs & Tesla S2050 & Tesla K20X\\\hline
                 1 &        0.48 &       1.13\\\hline
\end{tabular}
\end{center}
\caption{{\small GTEPS obtained by running the 2D code on a Tesla S2050 (Fermi) and on a
Tesla K20X (Kepler) using an R-MAT graph with scale $21$ and edge factor $16$.}}
\label{tab:tesla_kepler}
\end{table}

\begin{table*}[ht]
\centering
{\fontsize{10}{11}\selectfont
\resizebox{0.9\textwidth}{!}{
\begin{tabular}{| c | c | c | c | c | c | c |}
\hline
Data Set Name    & Vertices & Edges      & Scale     & EF         & GPUs & GTEPS       \\\hline
com-LiveJournal  & 3997962  & 34681189   & $\sim22$  & $\sim 9$   & 2    & 0.77 (0.43) \\
soc-LiveJournal1 & 4847571  & 68993773   & $\sim22$  & $\sim 14$  & 2    & 1.25 (0.47) \\
com-Orkut        & 3072441  & 117185083  & $\sim22$  & $\sim 38$  & 4    & 2.67 (1.43) \\
com-Friendster   & 65608366 & 1806067135 & $\sim25$  & $\sim 27$  & 64   & 15.68 (5.55)\\\hline
\end{tabular}
}}
\caption{{\small BFS performance of the 2D code with real-world graphs from the Stanford
collection. For each graph are reported the number of vertices, of edges, and
the (approximate) scale and edge factor (to facilitate the comparison with RMAT
graphs). Each traversal has been performed 64 times starting from random
vertices, with the specified number of GPUs. Performance reported are the
harmonic means of the TEPS measured in the traversals. Within parenthesis are
reported the performance figures of the original code.}}
\label{table:realgraph}
\end{table*}

Figure \ref{fig:comp_comm} reports the compute and transfer times of the code.
The compute time is the aggregate time required by the frontier expansion and
update phases to process the data received in the communication phases.
Keeping the problem size {\em per} processor fixed, the amount of data received by any
process (remote frontiers and discovered vertices) grows with the row and column
sizes of the processor grid.  For that reason, when the number of
processors grows also the amount of data to be exchanged increases,
and so it does the execution time.
Transfer time includes the time spent in the expand and fold exchanges and in
the {\em allreduce} operation at the end of the BFS loop. Up to 2048 GPUs, data
transfers require less than half of the total BFS time. With 4096 GPUs the
communications become dominant and take almost $60\%$ of the total search time.
Above 4096 GPUs, we do not expect the code to scale anymore but we did not test
it directly.

In Figure \ref{fig:breakdown} it is reported the timing breakdown of the four
steps performed in the BFS round: expand exchange, frontier expansion, fold
exchange and frontier update. Times are summed across the BFS steps and
averaged across 64 operations. By increasing the number of processors, the time
required by the frontier expansion phase reduces and, as expected, the weight of
the communications increases, becoming dominant with 4096 GPUs. The time required
by the frontier update phase is always much smaller than the time spent in the
frontier expansion. With any configuration, it accounts for less than $10\%$ of
the total time required by the four steps.

In Figure \ref{fig:1D2D_comp} we report the comparison of computation and
communication times between our original code and the new one, both run on the
Piz Daint cluster.\rem{The 2D code outperforms the original code in both
communications and computations.} For what concerns the communications (right
plot), the advantage of 2D partitioning is apparent with any number of GPUs up
to 1024, where 2D transfers are almost eight times faster than those in the
original code. For what concerns the computations, with 1024 GPUs the 2D code is
${\sim}3.5$ times faster. This advantage is mainly due to the performance
improvement of the atomic operations implemented in the Kepler architecture.
Figure \ref{fig:1D2D_front_exp} shows a comparison of frontier expansion times
between our present code using atomics and our original code.  Starting from
$16$ GPUs, the frontier expansion performed with atomic operations is $2$ times
faster than our original approach supporting benign race conditions.

In Table \ref{tab:tesla_kepler}, we report the performance of the 2D code
obtained on a single GPU with both the Fermi and Kepler architectures. The code
runs more than twice as fast on the Kepler GPU.

Finally, Table \ref{table:realgraph} reports the results obtained traversing
real-world graphs obtained from the Stanford Large Network Dataset Collection
\cite{slndc}. Among them we selected undirected graphs with the highest number
of edges.

\section{Latest Communication optimization}\label{sec:compression}

The results presented in Section \ref{sec:results} (Figures
\ref{fig:comp_comm}, \ref{fig:breakdown}) show that the communication times
steadily grow  and eventually exceed the compute time using 4096 nodes. In
order to reduce communications cost we reduced the number of vertices exchanged
by using a bitmap to keep track of the visited status of both local and
non-local vertices.  To further reduce the size of the messages we decided to
fit to our code an approach proposed in \cite{Satish:2012, IBMSC2012}
consisting in using bitmaps also for data transfers. The idea is that when the
size of vertices lists to be sent exceeds, in bits, the number of indices local
to the receiving process, then it is more convenient to send a bitmap with the
bits corresponding to the outgoing vertices set.  This technique reduces the
communication times by limiting the data transmitted to a fixed amount whenever
the number of vertices to be transferred would grow over a given threshold.
Obviously, that happens in the most expensive steps of the visit.  Assuming
that the indices to be transferred are 32-bit words and that the size of the
subgraph per processor is fixed and equal to $SCALE=21$ we have that the bitmap
size is, regardless of the number of processors:

\[
size_{b} = \frac{1}{8}\frac{N}{RC} bytes = 256 kb
\]

\noindent
With respect to the number of vertices, the threshold $T$  over which it is
more convenient to use a bitmap is:

\[
T = \frac{size_{b}}{4} = 65536
\]

Whenever a list with more than $T$ vertices needs to be sent, just $256$Kb of
data are actually sent, independently of the size of the list (that can grow up
to $8$Mb in the example above).  This chance applies to both the frontier
expansion and fold phases.

In order to maximize the benefits, it is necessary to use the bitmaps only
when the number of vertices in the list exceeds the threshold $T$.

We implemented a simple mechanism to dynamically switch between the two
representations, using the most convenient alternative for the two transfer
procedures at each step of the visit.  It is worth noting that with R-MAT
graphs the switching point happens during the BFS step, between the expand and
the fold phases. This is in accordance to the analysis of the expansion and
contraction of the frontier reported by previous studies \cite{Merrill:2012,
Beamer:2012, BaderGPU2014}.

For what concerns the packing/unpacking of the vertices lists into/from
bitmaps, we used two different approaches. The packing has been implemented
into specialized versions of the frontier expansion and update kernels that
directly produce bitmaps instead of lists. For the unpacking, we developed
specific warp-centric kernels based on the Kepler's shuffle instructions. The
overhead introduced by these kernels is negligible with respect to the frontier
expansion/update kernels and largely compensated by the dramatic reduction of
the data-exchange times that it makes possible.

\begin{figure}
\begin{center}
\includegraphics[scale=0.5]{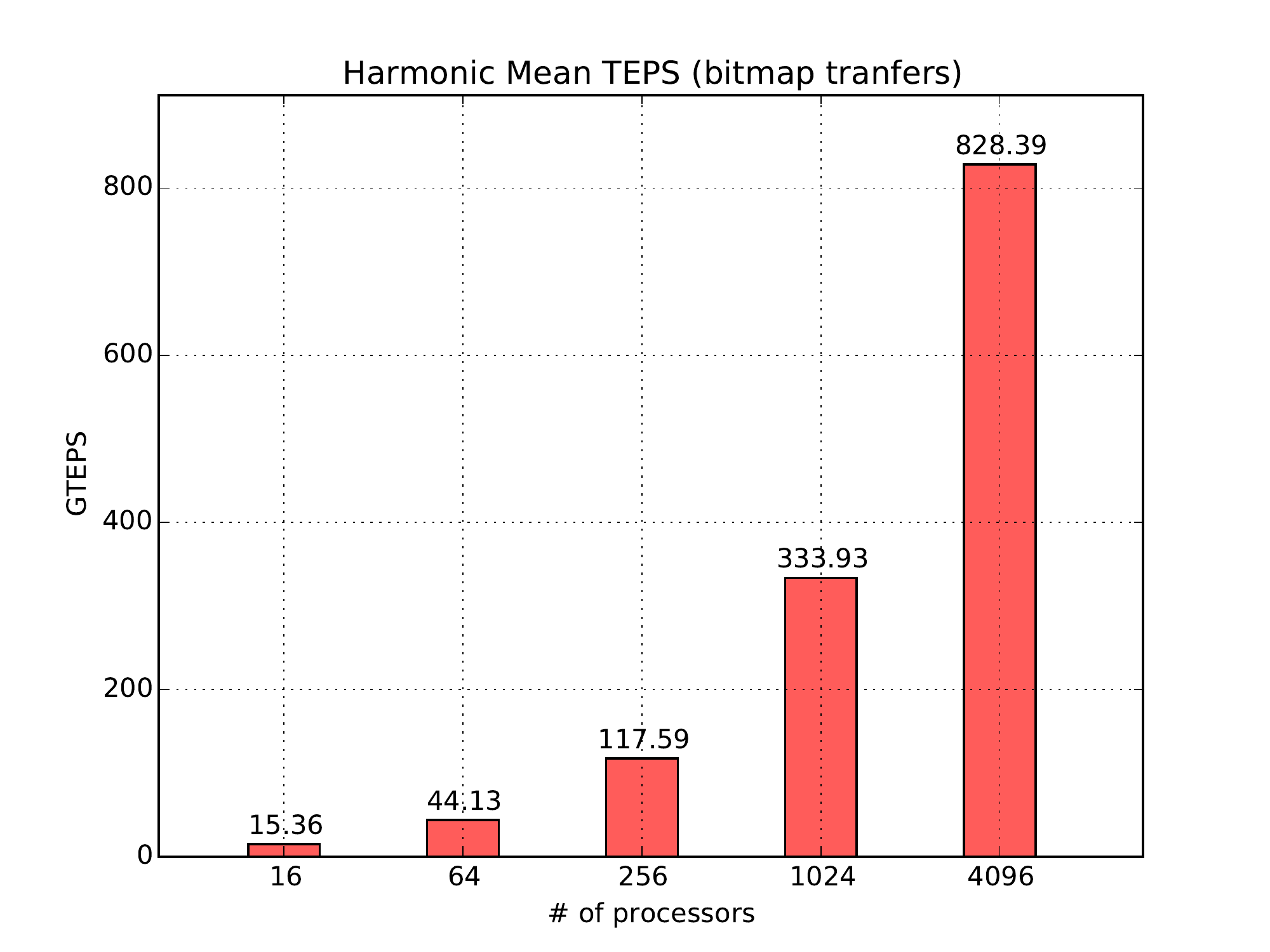}
\end{center}
\caption{Harmonic mean TEPS measured with a number of GPUs ranging from 1 to
4096 keeping the graph scale per processor fixed and taking advantage of the
bitmap optimization.}\label{fig:teps_bitmap}
\end{figure}

\begin{figure}
\begin{center}
\includegraphics[scale=0.5]{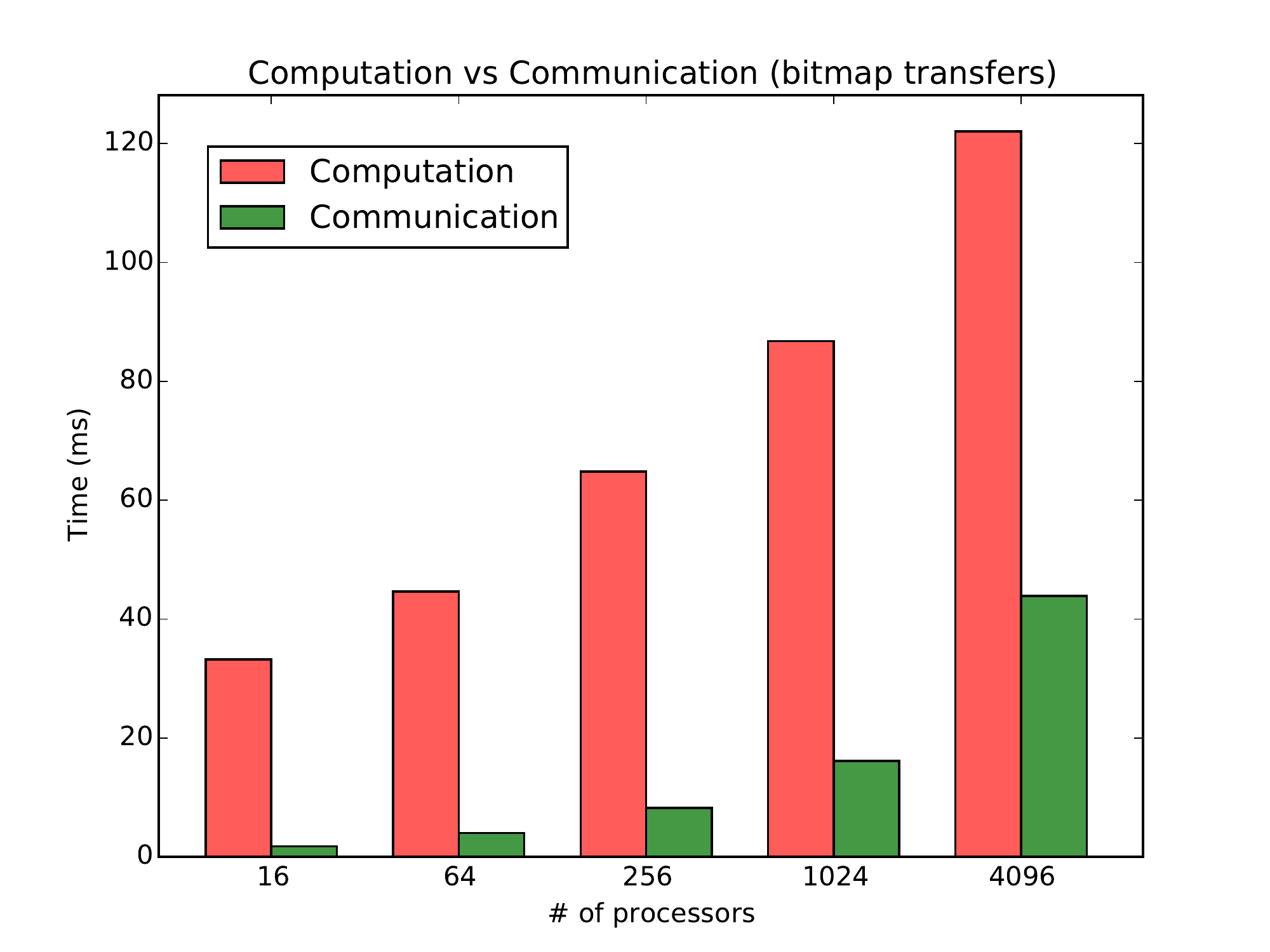}
\end{center}
\caption{Computation and transfer times (left and right bar, respectively) for
BFS operations performed with variable number GPUs on R-MAT graphs taking
advantage of the bitmap optimization.}
\label{fig:comp_vs_comm_bitmap}
\end{figure}

The impact of this optimization on the overall performance is remarkable and
leads to a net 2x speed-up in the number of TEPS as shown in Figure
\ref{fig:teps_bitmap}. Figure \ref{fig:comp_vs_comm_bitmap} reports the
aggregated computation and communication times of the code using the bitmaps.
The drop of communication times from ${\sim}200$ms to ${\sim}40$ms with 4096
GPUs is an indication that the code may scale even on larger GPU clusters.

\section{Related works\label{sec:related_works}}

In recent years high performance implementations of graph algorithms have
attracted much attention. Several works tackled the issues related to both
shared memory and distributed memory systems.

On the single GPU different solutions have been proposed to address workload
imbalance among threads.  The naive assignment, in which each thread is assigned
to one element of the BFS queue, may determine a dramatic unbalance and poor
performances \cite{Mastro2013}.  It is also possible to assign one thread to
each vertex of the graph but, as showed by Harish {\em et al.}
\cite{Harish:2007kx}, the overhead of having a large number of unutilized
threads results in poor performances.  To solve this problem, Hong {\em et al.}
\cite{Hong:2010, Hong:CPU}  proposed a warp centric programming model.  In their
implementation each warp is responsible of a subset of the vertices in the BFS
queue.  Another solution has been proposed in the work of Merrill {\em et al.}
\cite{Merrill:2012}. They assigned a chunk of data to a CTA (a CUDA block). The
CTA works in parallel to inspect the vertices in its chunk.

We devised an original data mapping described in \cite{Mastro2013}. Then, to
reduce the work, we used an integer map to keep track of visited vertices.
Agarwal {\em et al.} \cite{Agarwal:2010}, first introduced this optimization
using a bitmap that has been used in almost all subsequent works.

On distributed memory systems many recent works rely on a linear algebra based
representation of graph algorithms \cite{Buluc:2011, UenoGPU, IBMSC2012,
Beamer_Distrib:2013}.

Ueno {\em et al.} \cite{UenoGPU} presented a hybrid CPU-GPU implementation of
the Graph500 benchmark, using the 2D partitioning proposed in \cite{Yoo:2005}.
Their implementation uses the technique introduced by Merrill {\em et al.}
\cite{Merrill:2012} to create the edge frontier and resort to a novel
compression technique to shrink the size of messages. They also implemented a
sophisticated method to overlap communication and computation in order to
reduce the working memory size of the GPUs.

A completely different algorithm that uses a direction-optimizing approach
has been proposed by Beamer {\em et. al} \cite{Beamer:2012} and extended in
\cite{Beamer_Distrib:2013} for a cluster of CPUs. The direction-optimizing
method switches between the {\em top-down} and the {\em bottom-up} traversal.
The {\em bottom-up} search dramatically reduces the number of traversed edges
during the most expensive computational levels of the BFS by searching the
parent in the frontier starting from the sub-set of unvisited vertices. The
parent search implies a serialization in order to minimize the required work.
However, on shared memory systems (including single GPUs  \cite{DanZouGPU2013,
BaderGPU2014}) the BFS performance increases significantly.

A recent work \cite{Petrini2014} demonstrates the chance of having an effective
implementation of a distributed direction-optimizing approach on the BlueGene/P
by using a 1D partitioning.  That partitioning simplifies the parallelization
of the bottom-up algorithm but it may require a significant increase in the number of
communications.  Their results show that the combination of the underlaying
architecture and the SPI interface is well suited to the purpose.  The authors
report that replacing the SPI with MPI incurs a loss of performance by a factor
of nearly 5 although the MPI-based implementation cannot be considered
optimal. This suggests that the scalability of a distributed implementation may
be worse on different network architectures.

The work presented in \cite{Hiragushi:2013} achieves an outstanding speed-up
for the implementation of a direction-optimizing BFS on a single GPU,
nearly six times faster than more recent implementations
\cite{BaderGPU2014, DanZouGPU2013}.
However, it also shows a very poor scalability on a multi-GPUs system.

It should be also taken into account that the bottom-up approach may not be a
viable option if an actual traversal of all the edges in a connected component
is required.

Satish {\em et al.} \cite{Satish:2012} implemented a distributed BFS with 1D
partitioning that shows remarkable scaling properties up to 1024 nodes.  They
devised a technique to delay the exchange of predecessors. We implemented the
same technique independently, as reported in \cite{Bernaschi2014}, and use it
in the present work.

\vspace{0.5cm}
Compared with state-of-the-art implementations on distributed architectures,
our implementation on 4096 GPUs is 2.6 times faster than the distributed GPU
implementation in \cite{UenoGPU} that uses 4096 GPUs and 1366 CPUs, 3.4 times
faster than the distributed bottom-up in \cite{Beamer_Distrib:2013} that uses
up to 115000 cores and 3.3 times faster than the 2D implementation in
\cite{IBMSC2012} that uses 4096 BlueGene/Q nodes.  To the best of our knowledge,
the work in \cite{Petrini2014} achieves the best performance.

\section{Conclusions\label{sec:conclusions}}

We presented the performance results of our new parallel code for distributed
BFS operations on large graphs. The code employs a 2D partitioning of the
adjacency matrix for efficient communications and uses CUDA to accelerate local
computations. The computational core is based on our previous work on GPU graph
processing and is characterized by optimal load balancing among GPU threads,
taking advantage of the efficient atomic operations of the Kepler architecture.

We further enhanced the code by using a bitmap to reduce the data exchanged
among processors during the most expensive BFS steps.  This optimization,
improved the performances up to a factor 2 with 4096 GPUs.

The result is a code that shows good scalability up to $4096$ Nvidia K20X GPUs,
visiting $830$ billion edges per second of an R-MAT graph with $2^{33}$
vertices and ${\sim}280$ billions of directed edges.  We compared the
performances of the new code with those of the original one, which relied on a
combination of parallel primitives in place of atomic operations, on the same
cluster of GPUs. The 2D code is up to eight times faster on R-MAT graphs of the
same size.

\section*{Acknowledgments}
We thank Giancarlo Carbone, Massimiliano Fatica, Davide Rossetti and Flavio
Vella for very useful discussions.

\bibliographystyle{unsrt}
\bibliography{graph500_201412}{}

\begin{thebibliography}{10}
\bibitem{2DAlgOpt}
In order to reduce the amount of exchanged data, in the fold phase we send only
  destination vertices, instead of edges. moreover, to avoid sending more than
  once the same vertex to its owner, we use a bitmap to keep track of local
  vertices that have been visited.
\bibitem{BinSearchAcc}
Since consecutive threads search for consecutive values, sequences of threads
  mapped onto the same adjacency lists access the same memory locations during
  the binary search. different sequences share an initial part of the search
  path whose length depends on their relative distance. as a consequence, the
  memory access pattern is quite efficient since threads in each sequence
  perform broadcast memory accesses that are served with single memory
  transactions, one for each search jump.
\end{thebibliography}
\end{document}